# Herausforderung Peer Instruction
## Bemühungen um einen doppelten Konzeptwandel in der Lehre
*von Isabel Braun (Hochschule Karlsruhe)*

Eine Veranstaltung mit Peer Instruction (nachfolgend PI) erkennt man vor allem an einem Merkmal: dem ständigen auf und ab der Lautstärke. Mal brodeln duzende parallele und teilweise sehr angeregte Diskussionen, dann ist wieder minutenlang absolute Stille. Die Stimme des Dozenten hört man weit seltener als in einer Vorlesung, und oft ist sie nur eine unter vielen. Wie Lehren und Lernen in so einem Ambiente gelingen kann haben bereits viele Lehrende aus den unterschiedlichsten Fachrichtungen erprobt (MAZUR 2009, DANCY & HENDERSON 2009).

Eine in PI gestaltete Veranstaltung greift in besonderem Maße auf Begriffe und Konzepte eines Faches zu und verbessert somit eventuell die Chance, Fehlvorstellungen[1], wie die im vorherigen Beitrag dieses Bandes von KAUTZ am Beispiel der ingenieurwissenschaftlichen Grundlagen vorgestellten, im Sinne des Faches zu transformieren. Studien, wie HAKE (1998) oder SMITH et al. (2011), geben zudem Hinweise, dass Peer Instruction für konzeptionelles Lernen „funktionieren" kann (dort *interactive engagement*), dennoch ist wie bei jeder Methode Vorsicht vor einer rein Kochrezept-artigen Anwendung geboten. In diesem Text werden daher neben der Methode auch einige Hintergrundüberlegungen vorgestellt um PI mit Leben zu füllen und mit Gewinn für alle Beteiligten einsetzen zu können.

Den Kern von PI bildet themenbezogene Kommunikation und Feedback. Feedback sowohl an den Lehrenden, der sonst eventuell nur das anfeuernde Nicken einer Minderheit als Indikator dafür hat, ob sein enthusiastischer Vortrag irgendeine Resonanz im Raum gefunden hat, Feedback aber auch an den einzelnen Lernenden, der einiges über den eigenen Kenntnisstand und auch den Vergleich zu jenem seiner Kommilitonen erfährt. Dabei ist das Feedback ehrlicher als eine Selbsteinschätzung oder auch als Rechenaufgaben es oft leisten können. Wenn Lehrende wie Lernende so eingefahrene Rollen verlassen, kann es intensiv oder anstrengend werden, Neugierde wecken und vor allem auch Spaß machen[2].

## 1. Einführung

### 1.1 Rahmen

Wie zahlreiche Projekte von Bund und Ländern belegen, befassen sich Hochschulen derzeit vermehrt mit ihrer eigenen didaktischen Qualität (z.B. www.qualitaetspakt-lehre.de). Dabei sind sie zum einen ihren Studierenden verpflichtet, zum anderen aber auch der Gesellschaft, welche im Zuge des „war for talents" (BEECHLER&WOODWARD 2009) einen Bedarf an qualifizierten Absolventen zeigt. Dieser Bedarf drückt sich auch in speziellen Curricula und letztlich in den Lernzielen einzelner Veranstaltungen aus.

---

[1] mit dem Begriff Fehlvorstellungen oder Fehlkonzept werden häufig (ab)wertende Assoziationen verbunden, welche außerhalb der Fachsprache eventuell nicht gültig sind. Daher verwenden andere Autoren auch Begriffe wie Alltagsvorstellung, alternative Vorstellung oder Präkonzept, insbesondere wenn die Richtigkeit der Expertenantwort anzweifelbar ist. Jeder der Begriffe hat jedoch einen eigenen Beiklang, so entstammen eventuell nicht alle Fehlvorstellungen dem Alltag oder sind nicht alle Präkonzepte als Vorstufe anzusehen. Im vorliegenden Text sind die Begriffe weitestgehend synonym zu verstehen.

[2] Hinweis: einige der in diesem Aufsatz ausgeführten Gedanken wurden von der Autorin bereits behandelt in (BRAUN 2014). Die weitere Ausarbeitung wurde aus Mitteln des Bundesministeriums für Bildung und Forschung unter dem Förderkennzeichnen 01PL11014 gefördert. Die Verantwortung für den Inhalt dieser Veröffentlichung liegt beim Autor.

Die gesellschaftlichen Rahmenbedingungen fordern zudem von einer Hochschule, Studierende mit heterogenen Zugangswegen und daher auch stark unterschiedlichen Vorerfahrungen und Eingangsvoraussetzungen zu einem gemeinsamen Lernziel zu führen. Für eine technisch orientierte Hochschule für Angewandte Wissenschaften ist dieses Ziel beispielsweise ein „kompetenter Ingenieur", welcher in seinem Arbeitsleben über ausreichend Kenntnisse verfügt, diese aber auch kreativ und verantwortungsvoll einzusetzen vermag. Der Weg zwischen Ausgangszustand und Ziel wird von verschiedenen Lehrenden mit unterschiedlichen Schwerpunkten begleitet, daher werden aufeinander aufbauende Curricula entwickelt, um die Lernveranstaltungen untereinander abzustimmen.

Die (Lehr-)Praxis bemüht sich um gangbare Wege und Methoden, um diese Anforderungen zu erfüllen. Das Format der Wissenspräsentation in Vorlesungen stösst dabei jedoch oft an seine Grenzen, wenn über Wissen hinausgehende Kompetenzen oder auch „nur Verständnis" angestrebt werden. Dieser Umstand kann lerntheoretisch begründet werden, daher kann es lohnend sein, alternative Methoden auch aus theoretischer Sicht auf diese Grenzen hin zu untersuchen. Nach einer Einführung in PI und dessen Verknüpfung mit anderen konzeptionell verbundenen Lehransätzen in den folgenden Abschnitten versucht das zweite Kapitel daher, die Wirkungsweise von PI aus einem konstruktivistisch motivierten Blickwinkel zu beleuchten und dabei besonders jene Punkte zu beachten, die bei rein frontal belehrenden Vermittlungsversuchen zu kurz zu kommen scheinen.

## *1.2 Ablauf einer Peer Instruction-Sequenz*

Der Ablauf von Peer Instruction ist überschaubar (vgl. MAZUR, 1997): Lehrende wählen im Vorfeld Fragen aus ihrem bestehenden Repertoire aus oder konstruieren neue. Idealerweise sind dies nicht (nur) reine Wissensfragen[3], sondern solche mit denen sie hoffen, verbreitete Fehlvorstellungen aufdecken zu können. Der Dozent entwickelt auf diesem Wege eine Abfolge von Denkfragen[4] („ConcepTests", siehe MAZUR, 1997), um das Verständnis eines einzelnen Konzeptes zu testen. Parallel dazu fällt die Entscheidung für ein Feedbackmedium, mit dem die Antworten *aller* schnell erfasst werden können. Wählt man Fragen, die als Multiple-Choice beantwortbar sind, kommen z.B. Abstimmungen mittels hochgehobener Finger einer Hand oder mit gut lesbaren Zahlen bedruckter Zettel („Flashcards") in Betracht. Falsche Antwortoptionen werden am besten aus gängigen Fehlvorstellungen, Alltagsvorstellungen oder aus zuvor gesammelten offenen Antworten erstellt. Zur genaueren Auswertung können sog. Audience Response Systeme (z.B. „Clicker" oder mit mobilen Endgeräten bediente Web-Anwendungen) verwendet werden. Einige dieser technologischen Lösungen lassen auch Fragen mit offenen Antworten zu (z.B. „learning catalytics", http://learningcatalytics.com).

---

[3] solche, mit denen Lernende zeigen, dass sie einen bestimmten Sachverhalt kennen und im gleichen Kontext wie dem, in dem der Sachverhalt gelehrt wurde, widergeben können.

[4] also solche, welche z.B. durch einen Transfer auf konkrete Phänomene versuchen, das Verständnis eines in der Fachkultur bestehenden Konzeptes abzufragen. Ein typisches Beispiel: Statt der Berechnung von Strömen in einem Schaltkreis wird gefragt, ob eingefügte Glühbirnen beim Umlegen eines zusätzlichen Schalters heller oder dunkler würden. Da die Helligkeit der Glühbirne vom Strom massgeblich bestimmt wird, sollten Personen, die den Strom in einem Schaltkreis berechnen können, auch das Helligkeitsverhalten abschätzen können. Aufgaben wie diese füllen einen großen Teil des Buches (MAZUR 1997).

In der Umsetzung werden die Fragen in der Vorlesung präsentiert und den Studierenden wird etwas Zeit zur individuellen Beantwortung gegeben. Dann wirft der Lehrende einen Blick auf die Antworten und entscheidet sich für eine der folgenden Optionen:

- Das Thema abschließen: Wenn die überwiegende Mehrheit die im Sinne des Experten richtige Antwort gefunden hat, geht der Dozent zum nächsten Thema über. Es empfiehlt sich, zuvor exemplarisch nach der Argumentation für diese Antwort zu fragen um auszuschließen, dass diese aus den falschen Gründen einleuchtend erscheint.
- Weiteren Input anbieten: Wenn kaum jemand die richtige Antwort gewählt hat, sollten zunächst Missverständnisse in der Aufgabenstellung ausgeräumt werden. Es kann aber auch notwendig sein, zunächst noch einmal Grundlagen zu wiederholen.
- Das Thema für die Partnerdiskussion öffnen: Bei einer mittleren Anzahl an richtigen Antworten, typischerweise zwischen 30% und 70%, kommt das eigentliche Kernstück von Peer Instruction zum Tragen: Die Diskussion mit einem Kommilitonen oder einer Kommilitonin, der oder die sich für eine andere Antwort entschieden hat. Die Studierenden werden aufgefordert eine Meinung zu identifizieren, die von der ihren abweicht, und diese Position zu widerlegen oder ihre eigene zu begründen. Jetzt schnellt die Lautstärke nach oben und die Aufmerksamkeit wendet sich vom Lehrenden ab, welcher nun die Freiheit hat, sich unter die Diskutierenden zu mischen und deren Argumente zu verfolgen, in „ruhigeren Regionen" die Diskussion durch Auffinden neuer Diskussionspartner für stille Studierende anzustoßen oder sich direkt mit Fragen oder Impulsen auf einzelne Teilnehmer einzulassen.
- Im Anschluss an die Diskussion erfolgt eine zweite Abstimmung, deren Ergebnis der Lehrende zu einer neuen Richtungsentscheidung nutzt. Wie beim direkten Abschluss sollte die Argumentation noch im Plenum zusammengeführt werden, wenn man sich für ein Fortschreiten im Thema entscheidet. Diese Plenums-Diskussion kann auch genutzt werden, um gemeinsam neue Impulse zu entwickeln, sollten die Ergebnisse noch immer stark von der Expertenmeinung abweichen. Auch eine neue Diskussion mit anderen Partnern oder die Bearbeitung einer eingeschobenen, einfacheren Frage zum gleichen Thema kann hilfreich sein.

Auf die oben beschriebene Vorbereitung und die in Abb. 1 noch einmal schematisch zusammengefasste Durchführung hin, kann nach der Veranstaltung beim Lehrenden eine Phase der Reflexion stattfinden: Welche Konzepte haben den Studierenden am meisten Schwierigkeiten bereitet, welche Antworten oder Argumentationsketten haben mich überrascht, wo und wie sollte die nächste Veranstaltung anknüpfen? Mit diesem Reflexionsschritt entwickelt sich auch das Verständnis für im eigenen Fach verbreitete Studierendenvorstellungen stetig weiter, sowohl in der Person des Lehrenden als auch möglicherweise in der fachdidaktischen Gemeinschaft.

Entwickelt wurde diese Methode von Eric Mazur, Physik-Professor an der Harvard Universität, als er erkannte, dass seine Studierenden zwar komplizierte Rechenaufgaben lösen konnten, aber bei einfacher erscheinenden, eher qualitativen „konzeptionellen" Aufgaben scheinbar zielsicher die falschen Antworten gaben. Getrieben von der Frage, warum dies auch nach klaren Herleitungen und Erklärungen den meisten Studierenden so schwer falle, bemerkte er, dass viele Konzepte von einem Kommilitonen, der sie gerade erst verstanden hatte, im Partnergespräch erfolgreicher erklärt werden konnten als von ihm als Fachexperten (Expertenproblem oder „expert blind spot", siehe auch HINDS 1999). Daraus entwickelte Mazur PI als Versuch, im sokratischen Sinne durch Fragen zu lehren (MAZUR, 1997).

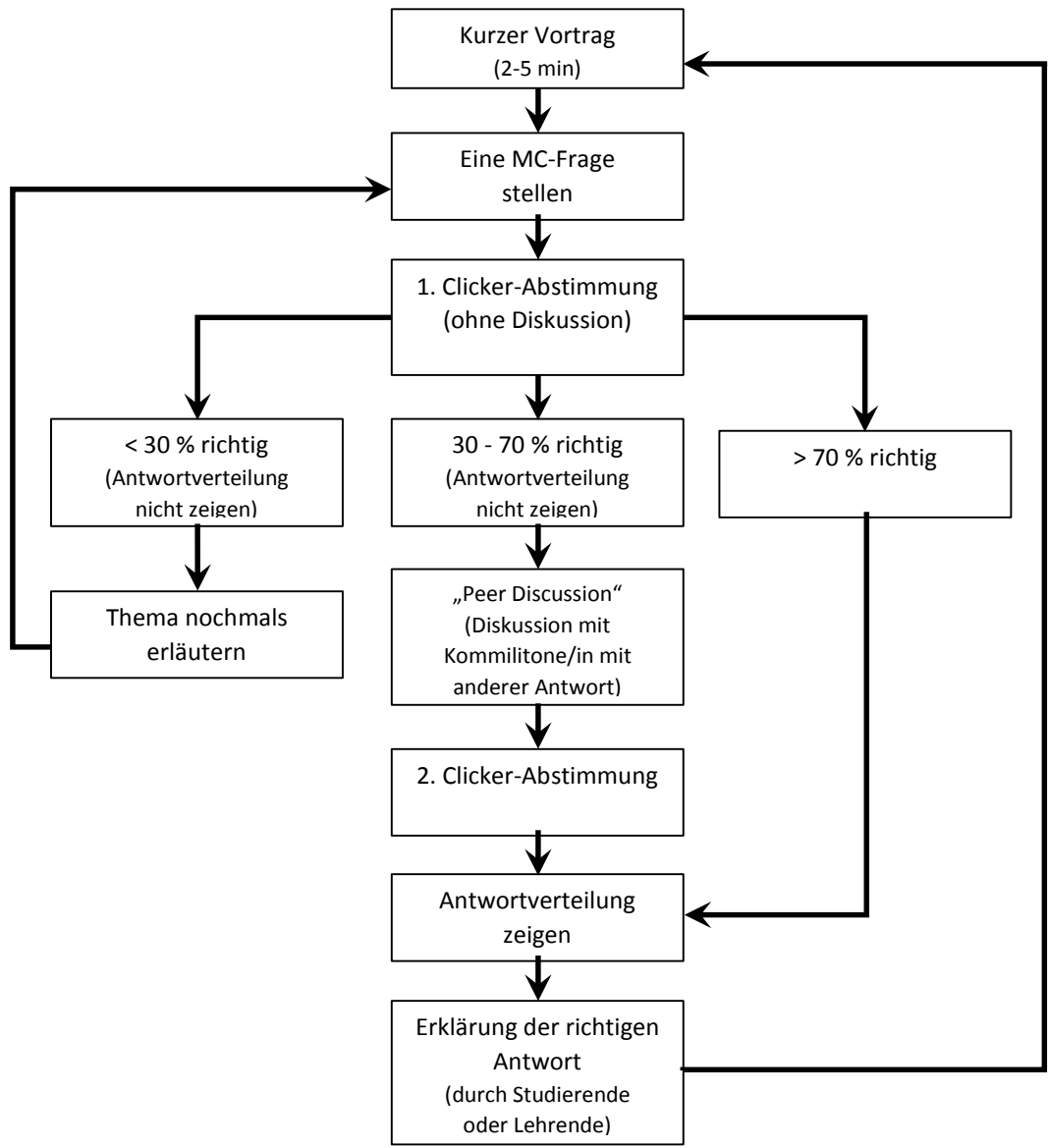

***Abbildung 1***: *Ablauf Peer Instruction nach* MAZUR *(2013), überarbeitet von Mikko Vasko*

### *1.3 Schnittstellen zu anderen Methoden: Inverted Classroom und Just-in-Time Teaching*

Ein häufig gegen die Einführung vertiefender Methoden angeführtes Argument ist der zusätzliche Zeitbedarf. Natürlich dauert eine Peer Instruction Sequenz länger als die alleinige frontale Präsentation der gleichen Inhalte. Insgesamt können Studierende dadurch aber sogar Zeit gewinnen, da er bei der Vertiefung Unterstützung hatte. Für den Lehrenden ist es aber Vorlesungszeit, die nicht mehr für die Präsentation neuen Stoffes zur Verfügung steht. Will man mehr als nur gelegentliche Zwischenfragen stellen, kann eine Neubetrachtung des Lehrablaufes Abhilfe schaffen. Mazur (1997) schlägt vor, die Wissenspräsentation der gemeinsamen Veranstaltung in Form von Leseaufträgen vorzulagern. Eine modernere Version dieser Idee ist das *Inverted Classroom Modell* (ICM) (vgl. BISHOP & VERLEGER, 2013). Auch im ICM wird die Präsentation des Stoffes der Veranstaltung vorgelagert und den Studierenden zum Selbststudium zur Verfügung gestellt. Oft geschieht dies in Form von Videosequenzen, eventuell kombiniert mit anderen Materialien wie Leitfragen, Lehrbuchauszügen oder Lückentexten. Im einfachsten Fall sind es auch Leseaufträge und Aufzeichnungen von früheren Vorlesungen, die Abgrenzung des Begriffs ist jedoch in der aktuellen Literatur noch nicht scharf: einige Autoren halten die Videos für den zentralen Aspekt von ICM (BISHOP & VERLEGER, 2013), andere verwenden Texte, wodurch auch vorgelagerte Leseaufträge, wie in MAZUR (1997) im weitesten Sinne eine Spielart oder zumindest eine Vorstufe des ICM darstellen.

Um den Anreiz zu erhöhen, diese Vorbereitung rechtzeitig durchzuführen, kann das dadurch erreichte Vorwissen kurz abgeprüft werden. MAZUR (1997) bietet daher eine Reihe von einfacheren Wissensfragen an, die zu Beginn der Veranstaltung von seinen Studierenden beantwortet werden mussten. Diese Kurztests können den Studierenden aber auch online als Lernkontrolle zu Verfügung stellen. Entsprechend werden beim *Just-in-Time-Teaching* (JiTT) Konzept aus dem Umfeld von Blended-Learning Veranstaltungen (NOVAK et al. 1999), solche Tests kurz (ein paar Stunden oder am Abend) vor der Veranstaltung fällig. Beim JiTT wird der Wert offen zu beantwortender Fragen betont insbesondere für den Lehrenden, da er mit ihnen die besten Einblicke in verbliebene Verständnisprobleme oder Missverständnisse erhält. Hierzu erhalten die Studierenden den Auftrag, am Ende eines Lernkontroll-Tests eine Frage zum Themengebiet zu stellen. Die führt mitunter zu sehr spannenden Einblicken in die Lernprozesse und Interessen der Studierenden. Der Dozent wertet den Test vor der Veranstaltung aus und stimmt seinen Unterricht auf den Bedarf (die evidenten inhaltliche Lücken) und das Interesse (die gestellten offenen Fragen) der Teilnehmenden ab. Im JiTT gestellte Fragen und daraus gewonnene Erkenntnisse zu bestehenden Verständnisproblemen können als Ausgangsbasis für neue PI-Fragen verwendet werden. Die Veranstaltung wird individueller und ändert sich von Kurs zu Kurs, auch für den Lehrenden bleibt es so abwechslungsreich.

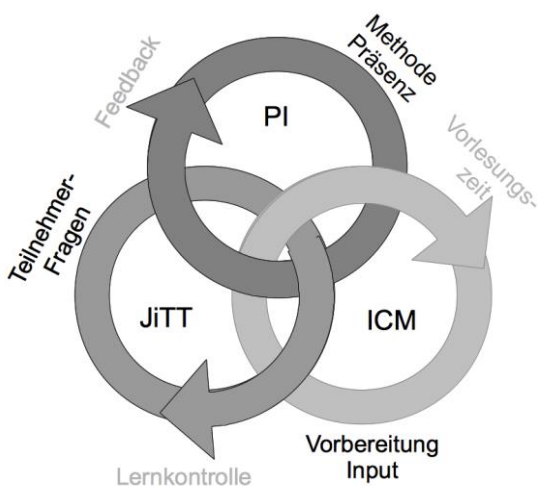

***Abbildung 2:*** *In der Kombination der Lehransätze PI, ICM und JiTT profitiert jeder Teilaspekt. Die dunkler gedruckten Stichworte verdeutlichen, was eine Methode aus dem vorgelagerten Kreis „mitnehmen" kann, in grau wird die Herausforderung aufgezeigt, welche von der nachfolgenden Instanz angegangen wird:*
*- JiTT „füttert" PI mit neuen Fragen der Teilnehmer, PI ermöglicht im Gegenzug weiteres Feedback an die Teilnehmer, z.B. zu den offenen Fragen.*
*- PI bietet sich als vertiefende Methode in der ICM Präsenzveranstaltung an. Will man PI einsetzen erhält man durch ICM Zeit in der Vorlesung.*
*- ICM kann als Vorbereitung von JiTT dienen. Die Lernkontrolle in JiTT kann wiederum im Gegenzug diese Vorbereitung motivieren und vertiefen.*

Als Einheit aus individueller Vorbereitung, beispielsweise über Videos (ICM), online Lernstandskontrolle und Interessensabfrage (JiTT) sowie daraus entwickelter verständnisorientierter Partnerdiskussion (PI) ergibt sich ein flexibles Blended-Learning Veranstaltungskonzept (siehe Abb. 2). Ob man nur gelegentlich einzelne Kontrollfragen einstreut oder die komplette Veranstaltung nach den Prinzipien des Inverted Classroom oder Just in Time Teaching umstellt, sowie die Ausgestaltung der Fragen und der Diskussionsphasen, beinhaltet viele Möglichkeiten zur Findung des eigenen Stils und zur Anpassung an die Gruppendynamik des Kurses. Die theoretischen Betrachtungen in Kapitel 2 stellen einen Versuch dar, Empfehlungen für einige dieser methodischen Entscheidungen abzuleiten.

## 2. Theoretische Betrachtung spezieller Aspekte der Umsetzung

Studierende haben zwar meist ein gemeinsames Studienfach aus eigenen Antrieb gewählt und sind so bezüglich ihrer Interessen homogener als beispielsweise eine Schulklasse, haben aber als (zumeist junge) Erwachsene im Gegenzug besonders in den ersten Semestern bereits sehr vielfältige Vorerfahrungen mit dem Themenbereich. Daher greife ich im Folgenden als lerntheoretische Grundlage Arbeiten der Erwachsenenbildung auf. Besondere Beachtung gilt dabei konstruktivistischen Ansätzen, da diese die Bedeutung der persönlichen Anschlussfähigkeit betonen.

Als Kernthese einer systemisch-konstruktivistisch verstandenen Erwachsenenbildung kann man den Satz „Erwachsene sind lernfähig, aber unbelehrbar" ansehen (SIEBERT 2012, S. 97). Diese Grundannahme betont die Unmöglichkeit sicherer Lehrrezepte und die Mitverantwortung der Lernenden an jeglichem Lernerfolg. Die Erkenntnisse konstruktivistischer Theorien stellen den direkten Wissenstransfer durch die pure Präsentation der Ergebnisse und Schlussfolgerungen anderer Personen in Frage (z.B. bei HOLZKAMP (1993) unter der Bezeichnung „Lehr-Lern-Kurzschluss"). Jeder Lernende konstruiert sein Wissen selbst und ist somit aktiver Gestalter des Lerngeschehens, nicht passiv belehrbares Objekt. Dieses Argument widerspricht deutlich der vorherrschenden Instruktionsdidaktik. Aus Perspektive der konstruktivistischen Theorie ist daher eine neue Betrachtung des Lehr-Lerngeschehens zu fordern. Aus jener sollten dann Kriterien herausgearbeitet werden, an denen man erkennen kann, ob solch eine Neubetrachtung gelungene Konsequenzen hatte. SIEBERT (2012, S. 60) fasst die Schwerpunkte einer konstruktivistische Didaktik zusammen durch den Ausdruck „Primat der Konstruktion".

Ich möchte die dort angesprochenen Punkte für diesen Text weiter auf die Aspekte der Individualität (in Vorerfahrungen, Bewertungen und Ankopplungsmöglichkeiten/Interessen – kurz: *„Bedeutung erhält, was passt"*), der Perturbation (*„Aufmerksamkeit erhält, was überrascht"*) und des sozialen Lernens (*„Richtig wird, was Konsens findet"*) verkürzen. Die nun folgende Beschäftigung mit erwarteten Wirkungsmechanismen soll schließlich bei der Entscheidung für oder gegen einzelne Aspekte des Peer Instruction unterstützen.

### *2.1 Individualität des Lernens und der Vorerfahrungen*

Die Annahme der Autopoiese von MATURANA & VARELA (1984) und die der Selbstreferenz oder Zirkularität des Lernenden betonen die Bedeutung individueller Vorerfahrungen. Lernende schöpfen aus ihrem jeweiligen Wissensschatz und Repertoire an Deutungsmustern. Gelingt es neuen Reizen vom Lernenden Beachtung zu erhalten, so werden diese von ihm interpretiert und im Allgemeinen dazu benutzt, das bestehende Repertoire an Deutungsmustern zu bestärken, nicht es in Frage zu stellen. Dies ist keine bewusste Entscheidung, sondern eine unumgängliche Antwort des Systems, welche eintritt noch bevor bewusste Entscheidungen möglich sind.

Gängige Fehlkonzepte werden daher in normaler Lehre kaum revidiert, auch wenn sie den Lehrenden eventuell aus vorherigen Klausuren vertraut sind und er/sie diese direkt anspricht. Werden PI Fragen speziell auf diese abgestimmt, kann jedoch immerhin ein Konflikt, entweder zu der Expertenmeinung oder einem Experiment, erzeugt werden, der beim Lernenden zu einer Perturbation (einer Irritation, ausgelöst durch einen Impuls von außen) führen kann.

Wenn Individuen neues Wissen nicht einfach aufnehmen können, sondern stets eine strukturelle Kopplung an bestehende Denkmuster als Voraussetzung für Lernen erforderlich ist, wird jede Information zunächst auf ihre Passung zur selbstempfundenen Wirklichkeit des Lernenden geprüft werden. Nur gangbare Alternativen werden akzeptiert. Die Viabilität (Gangbarkeit) trägt im Konstruktivismus Rolle und Funktion der Wahrheit (vgl. GLASERFELD, 1997, zit.n. SIEBERT, 2012, S. 54).

Im konkreten Kontext einer PI-Frage ist die Viabilität ein zentraler Aspekt, wird der „Wert" einer gewählten Antwort doch gemessen an der Robustheit der zu dieser Antwort führenden Argumentation. So wird trainiert, die Viabilität eines Konzeptes innerhalb der Logik der Lehrveranstaltung zu prüfen. Wendet der Argumentierende jedoch Alltagslogik zur Beantwortung der Frage an, wird die Bewertung der Viabilität seiner Lösung stark von der Passung an die Vorerfahrungen des Diskussionspartners abhängen, und das Team kommt eventuell nicht auf die vom Experten favorisierte Antwort. In diesem Fall empfiehlt sich, einen anderen Diskussionspartner zu finden.

'Studierende erhalten in Ihrem Studium Antworten auf Fragen, die sie sich nie gestellt hatten und Probleme, die sie nie empfunden haben' (vgl. VOSS, 2013), so das gängige Bild der MINT-Lehre. Nicht immer muss das aber so sein, denn Studierende bringen als erwachsene Lernende bereits viele Erfahrungen mit den meisten Lerninhalten mit. Die offene JiTT-Frage „Stellen Sie eine Frage zum behandelten Stoffgebiet" wird von einigen Studierenden zum Einbringen stets neuer Übertragungsmöglichkeiten oder Anwendungen genutzt, anhand derer sich die Inhalte oft ebenso gut behandeln lassen wie an den in Lehrbüchern abgedruckten Beispielen. So kann die Veranstaltung, verbunden mit etwas Mehraufwand für eventuelle Recherchen, an den konkreten Interessen und Ereignissen aus der Lebenswelt der Studierenden ansetzen und bleibt dabei stets aktuell und auch für den Dozenten ansprechend. Die Rückmeldung, dass studentische Anregungen ernst genommen und in der Vorlesung aufgegriffen werden, wirkt nach Erfahrungen der Autorin zudem förderlich auf das Lernklima im Allgemeinen und die Fragekultur im Speziellen.

JiTT und PI fördern auf diese Weise nicht nur die Anknüpfung des zuvor erworbenen Wissens an die Präsenzveranstaltung, sondern bieten auch die Möglichkeit zur Anknüpfung an die Lebenswelt und Deutungsmuster der Lernenden. Das Wissen wird passender und gleichzeitig relevanter.

### *2.2 Perturbation*

Da Lernende dazu neigen, neue Informationen in bestehende Wissensbestände derart einzubauen, dass nach Möglichkeit keine Irritation bestehender Ansichten eintritt, muss dem Individuum zunächst bewusst werden, dass sein bestehendes Wissen bei der Interpretation einer Situation nicht ausreichend ist. Wird durch einen solchen Widerspruch zwischen der eigenen Vorstellung und entweder einem Experiment oder der Expertenmeinung ein Lernbedarf wahrgenommen, ist das Subjekt irritiert oder perturbiert (hier synonym verwendet ohne wertende Konnotation, vgl. MATURANA & VERELA, 1984). Der Lernbedarf muss jedoch immer noch als nützlich oder relevant eingeschätzt werden, bevor weitere Lernaktivitäten eingeleitet werden. Für die subjektive Wahrnehmung einer Perturbation ist es daher hilfreich, den Konflikt des eigenen Wissensstandes an einem eventuell absurd einfach klingenden Beispiel zu erleben.

Ohne einen Moment der Perturbation gehen Studierende davon aus, dass Ihre Vorbereitung ausreichend war, zumal gut aufbereitete Lehrmaterialien oder Vorträge möglicherweise bereits eine Illusion des Wissens erzeugt haben. Bei einer besonders gut präsentierten und strukturierten klassischen Vorlesung wird diese Illusion eventuell sogar *verstärkt,* indem der eigene Kenntnisstand kurzzeitig durch das Verständnis des Dozenten überlagert wird (CARPENTER et al., 2013). Der Konflikt tritt erst im Nachhinein und in Abwesenheit des Experten auf, eventuell bei Übungsaufgaben oder gar erst in der Klausur. Auf diese Art kann eine zu glatte, perfektionierte Vorlesung kontraproduktiv wirken, da der Lernende die Inhalte als logisch präsentiert abnickt, ihm das Thema selbst in seiner vollen Komplexität und Widerspenstigkeit jedoch entschlüpft. Die einzige bleibende Spur im Lernenden kann so die Erinnerung an eine unterhaltsame, „gute" Vorlesung sein, jedoch ohne den Inhalt hinter der empfundenen Zufriedenheit später noch zu kennen.

In PI wird explizit Gelegenheit zu einer konstruktiven Irritation gegeben, und zwar insbesondere dann, wenn sich jeder individuell auf eine Antwortoption festlegen muss. Auch die einen PI-Durchgang abschließende Diskussion, in der Begründungen für gewählte Alternativen diskutiert werden, kann zu weiteren produktiven Konflikten mit den eigenen Vorstellungen führen.

Wird eine Perturbation erzeugt, sollte man beachten, dass diese ausreichend bearbeitet werden kann, damit der neue Weg nicht als Konflikt mit eigenen Überzeugungen verstanden und abgelehnt wird. Dies kann selbst in großen Kursen gelingen, da im Partnergespräch mit einem idealen Betreuungsverhältnis von 1:1 an der Ergründung, Widerlegung und Rekonstruktion von bestehenden Deutungen gearbeitet wird.

## *2.3 Soziales Lernen*

Lernen auf diese Art ist soziales Lernen. Wissen erhält seine Bedeutung in der Personalisierung des Arguments. Auch die wechselnde Quelle der zündenden oder überzeugenden Informationsfragmente verbessert deren Chance, behalten zu werden.

Die generelle Wirkungsweise von PI scheint mit dem Auftrag verknüpft zu sein, einen kognitiven Konflikt im Partnergespräch oder in einer Gruppe zu lösen. Präkonzepte werden dabei nicht umgangen, sondern bewusst aktiviert, allerdings (zunächst) nicht von der Lehrperson angesprochen. Die Verbalisierung eines Präkonzeptes durch einen Kommilitonen ist zwar sehr wahrscheinlich, wird aber typischerweise kritischer betrachtet als eine Erklärung des Lehrenden. Zugleich ist die eigene Rolle im Partnergespräch gleichberechtigter, und es ist leichter, sich aktiv einzubringen.

PI ist dann sowohl ein weiteres Feedbackinstrument („Classroom Assessment Technique") als auch ein Rahmen für soziales Lernen, denn die Konstruktion des übertragbaren Wissens geschieht größtenteils nicht beim Erstkontakt mit der Materie im Selbststudium, sondern im Gespräch mit einem Peer. Rückmeldungen von Clicker Ergebnissen fördern die Selbstwahrnehmung als Teil einer Gruppe und erhöhen die Fähigkeit der Selbsteinschätzung relativ zu dieser Gruppe. Leider verursachen konzeptionelle Denkfragen auch oft ein Gefühl der Unsicherheit, weil das vermeintlich Verstandene nicht sicher ausreicht, um allen potentiellen Transferfragen[5] gewappnet zu sein. Eine von mir beobachtete schützende Ausweichstrategie mancher Lernenden ist dann, gerade bei freiwilliger Vorbereitung: „Ich habe ja in der Vorbereitung noch nicht richtig gelernt, wenn ich direkt vor der Klausur mit dem echten Lernen anfange, werde ich das alles natürlich beantworten können".

In diesen Bereich fällt eine weitere Hypothese, die neben dem angesprochenen Expertenproblem das Lernen von Kommilitonen erfolgreicher machen könnte: die bessere Passung der Denkstrukturen und Deutungsmuster Studierender untereinander, welche oft aus einem ähnlichen Umfeld kommen und sich für ein ähnliches Ausbildungsziel interessieren. Die von einem Mitlernenden angebotene Argumentation kann so eventuell leichter vom Lernenden aufgenommen werden als jene des Lehrenden.

Es gibt Situationen, in denen man sich bei der Aufnahme eines Gedankenbogens noch nicht bewusst ist, wo er einen hinzuführen vermag, ein Effekt der von Heinrich von Kleist 1805 beschrieben wird im Aufsatz „Über die allmähliche Verfertigung der Gedanken beim Reden".  Die Peer Diskussion bietet reichlich Gelegenheit für solche Momente, nach denen die Urheberschaft der Idee kaum noch zu identifizieren ist. Auf diese Art entwickeltes Wissen ist dennoch aus dem Individuum heraus entstanden, also selbst konstruiert, hat aber seinen Ursprung zugleich in der Interaktion der Gruppe.

---

[5] bei einer Transferfrage wird der gelernte Sachverhalt entweder auf einen Anwendungskontext übertragen oder in einer anderen Repräsentation (z.B. Graph statt Formel, ausformulierter Text) benötigt als jener, in welcher er gelehrt wurde.

## 2.4 Kritische Fragen und praktische Hinweise

In der Umsetzung sieht die Autorin verschiedene Sollbruchstellen, die dem Leser nicht verheimlicht werden sollen. Ganz generell stellt sich z.B. auch folgende Frage: Ist es nicht ebenfalls versteckter Behaviorismus, zu glauben, Fragen würden bei allen Studierenden das gleiche auslösen? Es scheint, als wurde nur der Reiz „Wissen" durch den Reiz „Frage" ersetzt. Die Vermittlungsillusion besteht eventuell implizit weiter, wird aber nun auf Fragen abgelenkt. Der Widerspruch zur konstruktivistischen Sicht, in welcher letztlich immer der Lernende (mit-)bestimmt was gelernt wird, wird nicht aufgelöst.

Die Vielfalt an möglichen Umsetzungen von PI beschreiben TURPEN & FINKELSTEIN (2009). Aus den oben gezeigten Überlegungen ergeben sich Empfehlungen für die Umsetzung des Konzepts, die jedoch nicht als endgültige Wahrheit, Korsett oder Checkliste verstanden werden sollten, sondern als Anlass und Einstieg in eine individuelle Ausgestaltung dienen können. Sie setzen bei möglichen Bedenken an und greifen auf die theoretischen Überlegungen des vorangegangenen Kapitels zurück. Dazu werden im aktuellen Kapitel zwölf verschiedene, aber nicht unbedingt unabhängige Aspekte von PI beleuchtet, beginnend von eher systembedingten Herausforderungen zu potentiellen Schwierigkeiten auf Seite der Lernenden und schließend mit möglichen Bedenken und Vorbehalten der Lehrenden.

### (1) Selbstgesteuertes Lernen

Wird PI kombiniert mit ICM zur Wissensvermittlung, fordert dies ein hohes Mass an Selbstdisziplin und Lernkompetenz von den Studierenden, welche insbesondere in den ersten Semestern eventuell noch nicht ausreichend ausgeprägt sind. Dies taucht von Studierenden als Gegenargument auf („wenn man nicht vorbereitet ist, bringt die Veranstaltung nichts" - Studierender der Hochschule Karlsruhe 2013) und provoziert den Ruf nach klassischen Vorlesungen. Gleichzeitig ist selbstständiges Lernen (FRIEDRICH 2014) und das Formulieren eigener Fragen („Man macht keine Erfahrungen ohne die Aktivität des Fragens" (GADAMER 1975, S. 344)) von allen potenziellen Lernergebnissen wohl die wichtigste erreichbare Kompetenz. Bei einigen Lehrenden kann der Widerstand unzufriedener Studierenden dazu führen, dass sie deren Wunsch nachgeben und zur reinen Inhaltspräsentation zurückkehren. Daher sind Hilfen und Anleitungen zum selbstgesteuerten Lernen nützlich, welche z.B. im JiTT gegeben werden können. Über die Fragen wird inhaltlich, über den zeitlichen und formellen Rahmen methodisch steuernd eingegriffen, regelmäßige Bearbeitungsfristen fördern nach Erfahrungen der Autorin kontinuierliches semesterbegleitendes Lernen und beugen Prokrastination vor.

Dieser wenn auch überschaubare Zwang scheint schwer vereinbar mit konstruktivistischen Lehren, findet sich aber zumindest im Anforderungsprofil vieler Berufe wieder und ist unabdingbar angesichts der immer schneller werdenden Wissensveralterung, insbesondere im ingenieurwissenschaftlichen Bereich.

Neben dem Training im eigenständigen Erarbeiten neuer Inhalte wird durch PI eine Methode eintrainiert, mit der Lernende überprüfen können, ob das verstanden geglaubte Wissen in unterschiedlichen Situationen anwendbar und mit Messungen konsistent ist. Die allgemeine Kultur des Fragenstellens aus PI lässt sich auch auf neue Themen übertragen, und man gewöhnt sich an innere Fragen der Art „was bedeutet diese gelesene Aussage für...", eine für spätere wissenschaftliche Arbeit möglicherweise sehr fruchtbare Denkweise.

### (2) Gemeinsames Lernziel

Der Konflikt mit dem Konstruktivismus beginnt wohl bereits beim Ausgangspunkt, nämlich der Idee, ein gemeinsames Lernziel erreichen zu können. Nimmt man die Entkopplung von Gelehrtem und Gelerntem ernst, sollte dies prinzipiell unmöglich sein, da individuelle Lerner nur individuelle Lernergebnisse hervorbringen können. Bereits das Konzept der - dem Lerner aufgezwungenen - Lernziele kollidiert so mit dem selbstbestimmten, konstruktivistischen Lernen.

Warum trotzdem standardisierte Abschlüsse vergeben werden können scheint wundersam und liegt eventuell in der häufig beobachteten Praxis Studierender, kurz vor den Prüfungen die vom Prüfer gewünschten Formulierungen und Deutungen (auswendig) zu lernen. Dieses Auswendiglernen ist nur auf die Befriedigung der Ansprüche des Prüfers ausgelegt, nicht auf ein Aneignen der Inhalte. Erst wenn aufbauende Veranstaltungen auf die gleichen Kompetenzen zugreifen wollen und diese nicht mehr abrufbar sind, entsteht eine Perturbation mit subjektiv wahrgenommener Relevanz, die zur tieferen Bearbeitung aufruft. Aufeinander aufbauende Lehrveranstaltungen, idealerweise sogar bei unterschiedlichen Lehrenden, wären dann für das Lernergebnis am Ende eines Programms förderlicher als die extreme Modularisierung des Stoffes.

Auch in PI wird die Zieloffenheit konstruktivistischer Lehre nicht erreicht, nur in den Beispielen der Lernenden schimmert ein wenig thematische Freiheit durch. Echte Zieloffenheit ist jedoch aus curriculären Überlegungen wie in Kapitel 1.3 angesprochen selten erwünscht.

### (3) Richtigkeit der Expertenlösung

Ein weiterer möglicher Kritikpunkt ist die vermeintliche Richtigkeit der Expertenlösung. Auch wenn PI ambivalente Fragen zur offenen Diskussion unterstützt, gehen viele Lehrende implizit von der Idee aus, dass es dennoch immer genau eine richtige Lösung gibt. Dieses Bild der Wirklichkeit würde durch die Häufigkeit der Bewertungen durch Lehrende wahrscheinlich auf die Studierenden übergehen (wenn sie es nicht bereits von der Schule mitbringen), so dass es ihnen am Ende nur noch um die „richtige" Antwort geht, nicht mehr um die nötige Argumentationskette um zu ihr zu gelangen. Daher sollte der Peer-Diskussionsphase besondere Aufmerksamkeit gewidmet werden, z.B. indem der Lehrende die „Bühne" verlässt und sich als „Sparring-Partner" auf allzu abwartende Studierende einlässt. Es besteht sonst die Gefahr, dass die Diskussion als Zeitverschwendung wahrgenommen wird, wenn man eigentlich nur die richtige Antwort zur späteren Verwendung (z.B. in der Prüfung) notieren möchte.

Die Verwendung reiner Multiple-Choice Fragen im PI verschärft diesen Eindruck, weswegen viele Anwender in den USA versuchen, auf Systeme mit offenen Antworten überzugehen. Kurzfristig kann man es dadurch entschärfen, dass man wie in JiTT ursprünglich angeregt, zumindest in der online-Phase viel Wert auf Freitext-Fragen legt. Dies bedeutet zwar einen erhöhten Aufwand bei der Auswertung, ist aber als lohnend anzusehen.

Mit diesem Punkt kann einhergehen, dass Lernende durch die Tests in JiTT und PI eine vom Lehrenden möglicherweise unintendierte Stoffbewertung wahrnehmen: „was nicht im Unterricht gefragt wird ist auch nicht relevant" statt der ursprünglichen Idee „was bekannt ist wird nicht gefragt".

### (4) Des Rätsels Lösung

Es mag verlockend scheinen, den Ablauf (Abb. 1) noch weiter zu straffen, indem man auch die Auflösung im Plenum überspringt, wenn die Antworten fast durchgängig richtig waren.    Dieser Schritt ist riskant, z.B. kann die Antwort schon in der Vorbereitung überrascht haben und daher erinnert werden, obwohl sie eigentlich unverstanden ist. Meistens geben die Studierenden einem in diesem Fall zwar entsprechende Signale, es ist aber besser PI-Fragen generell nicht bei richtig/falsch belassen werden, sondern eine (kurze) Klärung in der Gruppe suchen, die auch die Argumentationskette beleuchtet. Schwierig ist dies wie im vorherigen Abschnitt auch, wenn die Lernenden einmal gewohnt sind, am Ende die richtige Antwort präsentiert zu bekommen, da die Teilnahme an der Diskussion zurückgehen und der PI-Prozess wie oben angesprochen zu einer lästigen Zeitverschwendung vor dem Input werden kann. Daher sollten Lehrende auch in der abschließenden Diskussion eher als Moderator auftreten denn als Wissensvermittler.

### (5) Sichtweise der Studierenden

Viele Studierende besuchen Vorlesungen mit einer bestimmten Erwartungshaltung, nämlich eine strukturierte Aufbereitung des Stoffgebiets zu erhalten. Diese Erwartung ist auch gerechtfertigt, denn gerade Strukturieren, Gewichten und Vernetzen von Wissensinhalten ist ein so persönlicher Prozess, dass die Sichtweise des Lehrenden sowohl als Beispiel dient als auch als Orientierung an den Schwerpunkten des Prüfenden. Der Wegfall einer solchen Orientierung erschwert nicht nur die Aneignung in der Selbstlernphase, mit der Auswahl eines Lehrbuchs kann auch die Ansicht vermittelt werden, dass dessen Logik und Struktur die einzig richtige ist. Der individuelle Charakter von persönlichen Erfahrungen mit dem Lerninhalt sowie die Begeisterung des Lehrenden für sein Thema kann unter den Tisch fallen, wenn es den Videos des ICM nicht gelingt, auch diese feineren Merkmale einer Kommunikation zu transportieren. Es können aber auch gerade hierfür Gelegenheiten entstehen, sei es durch die mittels ICM frei gewordene Zeit oder eine der Praxis und Leidenschaft des Lehrenden nahe stehende Auswahl an Fragen.

Lehre, die (gebräuchlichen Verfahren zur Lehrevaluation geschuldet) vor allem den Zufriedenheitserfolg bedient, wird Lernende selten aus ihrer gewohnten Praxis herausreißen. Eine radikale Umkehrung der Lernprozesse, wie sie ICM auslöst, birgt daher das Risiko, die Zufriedenheit der Teilnehmer zumindest kurzzeitig sogar zu reduzieren. Immerhin muss man zunächst selbst lernen, ohne belehrt zu werden (Widerspruch zur Lehr-Lern-Illusion nach HOLZKAMP, 1993), dies wiederum spricht die „gelernte Hilflosigkeit des Lerners" (ARNOLD 2009, S. 16) an und steht eventuell im Widerspruch zu früheren Lernerfahrungen.

Mangelnde Viabilität der Methode selbst kann insbesondere an Hochschulen zu einer generellen Ablehnung selbstgesteuerter Lehrmethoden führen, wenn Studierende als primäres Ziel „Erringen der Credits" erachten und nicht ausreichend vermittelt wird, dass Verständnis nicht als „Luxus" zu sehen, sondern tatsächlich „klausurrelevant" ist. Eventuell bedarf es daher einer umfänglichen und daher wahrscheinlich sehr unbequemen Anpassung der Prüfungsform.

Der erhoffte Gewinn für diesen Preis ist die Stärkung der Fähigkeit zur Selbstorganisation und anderer selbstschärfender Kompetenzen, welche die traditionelle Lehre nicht oder nur in geringerem Masse benötigt und daher auch weniger stark fördert. Nur leider werden diese Kompetenzen im ICM zwar vorausgesetzt aber nur sehr rudimentär vermittelt. Studierende, denen es noch nicht gelingt sich für die Vorbereitung zu motivieren, empfinden auch die Präsenz oft als Zeitverschwendung und sind noch schneller abgehängt als bei der klassischen Vorlesung. Eine daraus folgende, für PI gefährliche Deutung ist die von der Autorin mehrfach beobachtete Studierendenvorstellung „wenn ich mich vorbereitet hätte, hätte ich die Antwort auf die Transferaufgabe selbstverständlich gewusst, aber ich lerne lieber erst kurz vor der Klausur". Mit dieser Einstellung wird eine schnelle und bestehende Deutungsmuster sichernde Antwort auf die Irritation einer falschen Antwort gegeben und die Lerngelegenheit verstreicht ungenutzt. Eine Kontrolle der Vorleistung mit JiTT sollte hierfür hilfreich sein.

### (6) Soziale Anpassung

Besteht nicht die Gefahr, dass sich sozial schwächere Studierende von Meinungsführern überreden lassen, ohne dabei wirkliche Denkprozesse einzuschalten? Wenn Fragen als echte Transferfragen gestellt werden, kann auch die überzeugt vorgebrachte Antwort eines fortgeschrittenen Lerners manchmal fehlerhaft sein. Die Studierenden finden dies jedoch bald heraus und lernen aus dieser Erfahrung eventuell, jede Antwort, auch die von Experten, kritisch zu hinterfragen. Bei Fragen mit vielen verschiedenen gewählten Antworten (gelegentlich sieht man fast gleichverteilte Antworthäufigkeiten), kann es hilfreich sein, die Vielfalt des Ergebnisses rückzumelden. Studierende merken dadurch, dass alle mit dem Problem kämpfen und eigene Schwierigkeiten nicht die Ausnahme bilden, sondern die Regel. Außerdem hat man mit der eigenen Wahl immer auch falsche Antworten ausgeschlossen, deren Problem von anderen nicht bemerkt wurde. Es gibt so immer Anwesende, die von den eigenen Gedanken profitieren können. Die Autorin hatte in der bisherigen Anwendung nicht den Eindruck, dass sich Studierende nach derartigen Vorerfahrungen mit der kappen Antwort eines anderen zufrieden geben, sofern es nicht ihre eigene Meinung ist.

### (7) Wirkung auf Lehrende

Auch die Einstellungen des Lehrenden werden auf die Probe gestellt, denn bleibt man im lehrzentrierten, mechanistischen Weltbild, kann man verzweifeln wenn man die doch relativ geringen Lernzuwächse einer Erklärung misst. Wer messen will, muss auch mit dem Ergebnis zurechtkommen. Schreibt man sich die aus oben genannten Gründen unzufriedenen Studierenden zu oder leidet man unter beharrlich falschen Antworten, wird mitunter das Selbstbild als Lehrender gefährdet und die frühere Lehrmethode wahrscheinlich schnell wieder aufgenommen.

### (8) Sorge vor Falschem

Im Hörsaal werden nun potentiell fachlich falsche Aussagen gemacht und Fehlvorstellungen eventuell gerade dadurch aktiviert. Diese Vorstellung schreckt viele Lehrende ab, man sollte aber bedenken, dass auch bei einer fachlich korrekten Präsentation im Studierenden das falsche Muster aktiviert und abgespeichert werden (vgl. Crouch et al., 2004). In PI geschieht dies jedoch in der Diskussion mit einem Peer, nicht durch die Autorität des Lehrenden. Es bleibt daher zu hoffen, dass die Aussagen kritischer beleuchtet werden.

### (9) Vertrauen in den Weg zur Lösung

Selbst wenn in einem Paar keiner zunächst die richtige Antwort gewählt hatte, kann das gemeinsame Argumentieren zu dieser führen, wenn nämlich beide sich gegenseitig widerlegen. Da Widersprüche in der Diskussion aufgedeckt werden können, setzt sich die in der Logik der Lehrveranstaltung richtige Antwort oft auch dann durch, wenn kein Diskussionspartner sie ursprünglich gewählt hatte (Smith et al., 2009). Eine anschließende zweite Abstimmung zeigt meist einen deutlichen Zuwachs an richtigen Antworten.

### (10) Perturbation als Lernvoraussetzung

Warum doppelt abstimmen, ist das nicht Zeitverschwendung? Häufig wird die Abfrage individueller Ergebnisse in PI ausgelassen. Nicht nur um als Lehrender den Verlauf eventuell abzukürzen, auch um als Studierender eine eigene Meinung zu formen, ist diese erste Abstimmung sowie die Regieanweisung, zunächst noch nicht zu diskutieren, nützlich, denn es scheint für Lernende wesentlich ökonomischer und selbstförderlicher zu sein, die Auflösung abzuwarten und dann nachzuvollziehen. Diese Strategie weist aber nicht den gleichen Lernerfolg auf (vgl. CROUCH et al., 2004). Wie oben erörtert wird in konstruktivistischen Theorien die Perturbation der eigenen Denkmuster als zentrales Element für Lerngelegenheiten gesehen. Dafür muss diese aber vom Individuum zunächst einmal wahrgenommen und dann auch als relevant eingestuft werden. Ein Unterschied in den Items einer Multiple Choice Aufgabe ist womöglich leichter wahrzunehmen, als eine Schwachstelle in der eigenen Argumentationskette oder dem eigenen Weltbild. Bevor die Notwendigkeit einer Anpassung wahrgenommen ist, wird keine Anpassung des Weltbildes erfolgen, die Irritation ist also Voraussetzung für den Lernerfolg.

### (11) Bonus, Pflicht und Freiwilligkeit

Ein häufiger Streitpunkt ist die Frage der Freiwilligkeit von JiTT oder PI, bzw. der jeweiligen Bonierung entweder der reinen Teilnahme oder der richtigen Beantwortung aller Fragen. Für erkenntnissuchende, motivierte Studierende ist sicherlich die reine Freiwilligkeit des Angebots als Lernhilfe ideal, insbesondere wenn der Wert der Methode erkannt wird. Wenn jedoch die Vorbereitung von den Studierenden zugunsten anderer Verpflichtungen geopfert wird, verliert die Perturbation ihre Wirkung, da Probleme nun der mangelnde Vorbereitung und nicht mehr dem eigenen thematischen Verständnis attribuiert werden. In den ersten Semestern empfehle ich daher, die richtige Beantwortung von JiTT-Fragen mit einem Bonus zu versehen. PI-Fragen zu beantworten sehe ich jedoch als Teil des Lernprozesses und würde sie daher nicht bewerten. Professoren, die die PI-Teilnahme belohnen, berichten von zufällig erscheinenden Antwortverteilungen oder Studierenden, die mit den Abstimmungsgeräten von Kommilitonen in die Präsenzveranstaltung kommen - ein deutlicher Hinweis, dass die Methode selbst nicht mehr als Mehrwert wahrgenommen wird sondern nur noch der Bonus (als extrinsische Motivation) angestrebt wird.

### (12) Nach Fragen fragen

Für die hier aufgeführten Vorschläge gehen wir davon aus, dass Sie mehr als nur Zwischenfragen einsetzen möchten und dafür mittels JiTT oder ICM Ihre Veranstaltung entlasten wollen:

Die offene, wissenssuchende Frage in JiTT ist zentral für die Teilnehmerorientierung der Veranstaltung, auch wenn sie mehr Auswertungsaufwand bedeutet als vom Computer korrigierbare Rechenaufgaben oder Multiple-Choice Varianten. Eine themen- und fachübergreifende Möglichkeit hierfür ist die Aufgabe „Stellen Sie eine Frage zum behandelten Themengebiet". Sie ist mehrdeutig gestellt, daher werden einige Studierende nur fertige Fragen aus dem Buch oder aus bekannten Aufgaben wiedergeben, andere aber werden die Gelegenheit nutzen um eigene Lücken auszusprechen oder auf persönliche Interessen aus dem Umfeld des Themas hinzuweisen.
Der Umgang mit den Ergebnissen dieser Aufgabe muss unter allen Umständen wertschätzend erfolgen, um nicht zu oft die Antwort „Ich habe keine Fragen - alles verstanden" zu erhalten. Daher kann auch die Aussage „ich möchte keine Frage stellen" als richtig akzeptiert werden oder es kann eine Konvention mit der Bedeutung „meine Frage bitte nicht in die Gruppe tragen" vereinbart werden (z.B. der Antwort ein #-Zeichen voranstellen).

## 3. Fazit

Die Physikdidaktik für Schulen geht sehr sensibel mit Präkonzepten um („Das für den Physikunterricht wichtigste und zugleich am schwierigsten zu erreichende Ziel ist der Begriffswechsel", WIESNER et al., 2013, S. 89), plädiert aber häufig für den Aufbau einer fachlichen Begriffswelt neben der bestehenden Begrifflichkeit, da die interferierenden Alltagsvorstellungen einerseits für den Spezialfall unserer unmittelbaren Umwelt sinnvoll sein können, andererseits auch nicht sicher durch Fachkonzepte abgelöst werden können (WIESNER et al., 2013, S. 48). Für die Frage, ob solche multiplen Repräsentationen (Nebeneinander von Fehlkonzept und Fachkonzept) im Studium hingenommen werden sollten oder nicht doch ein Versuch zur Transformation von Präkonzepten erfolgen sollte, empfiehlt es sich, die Lernziele der Veranstaltung kritisch zu betrachten. Nach ERPENBECK (2010) können Kompetenzen überhaupt erst errungen, Wissen erst entscheidungsrelevant werden, wenn es in verinnerlichte Werte eingegangen ist. Eine sekundäre Repräsentation, welche nur aufgerufen wird, wenn man explizit zur fachlichen Argumentation aufgerufen wird, kann dies schwerlich leisten. KAHNEMANN (2011) argumentiert für die Unterscheidung zweier „Systeme": System 1 trifft schnelle Entscheidungen auf Basis unvollständiger Information, System 2 arbeitet analytisch. Bei der universitären Ausbildung sollen Experten auch in der Lage sein, im schnellen System unter Druck auf die Fachlogik zuzugreifen. Wissen muss also auf so viele Beispiele und Grenzfälle wie möglich übertragen werden, wenn es entscheidungsrelevant sein soll.

Dieses Übertragen auf Beispiele kann in PI erfolgen, wenn diese mit Blick auf jenen Aspekt formuliert werden. Die Autorin möchte daher dazu aufrufen, weiterhin die möglicherweise zum Scheitern verurteilte Herausforderung anzunehmen, Studierendenkonzepte im Sinne des Faches transformieren zu wollen.

Der Ausgangspunkt gemeinsamer Lernziele scheint mir wegen der Individualität der Lernprozesse prinzipiell nicht erreichbar. PI, insbesondere in Kombination mit JiTT enthält aber einige Aspekte, die dem Lernen aus konstruktivistischer Sicht förderlich sein sollten. Diese sind konkret die Erhebung und Berücksichtigung der Vorkenntnisse jedes Lernenden, die Einbindung eigener Interessen und Beispiele, sowie der Versuch, durch konzeptionelle Denkfragen Perturbationen zu erzeugen und diese in einem sozialen Kontext aufzulösen. Lernen wird als individueller und gleichzeitig sozialer Prozess realisiert, wobei das selbstständige Erarbeiten der Inhalte einen großen Platz einnimmt.

Auch die Kombination mit anderen Ansätzen kann im konstruktivistischen Sinne positiv wirken: Durch ICM erhält der Lehrende mehr Freiheit in der Gestaltung des Lernprozesses, da das lästige Abdecken von Materie nur, um es einmal gesagt zu haben, entfällt. Durch JiTT erhält der Lehrende eine neue Perspektive auf sein Fach: welche Inhalte sind wichtig, welche leicht oder besonders schwer verständlich, welche Beispiele greifen besser als das allgemeine Lehrbuch? Die Diskussionsphase innerhalb von PI ermöglicht Lehrenden wie Lernenden Einsichten in andere Perspektiven. Besonders wertvoll ist PI auch, wenn keine explizit richtige Antwort existiert, sondern Argumente und Gegenargumente gesammelt werden.

Eine mitunter auftretende Problematik ist die Schwierigkeit den Fokus trotz des verbreiteten Multiple-Choice Formates weg vom richtigen Endergebnis auf die Argumentationskette zu lenken. Auch sollte die Persönlichkeit des Lehrenden in den Inhalten und der Stoffstrukturierung erkennbar sein, da dies den absoluten Charakter der Lehrinhalte relativiert. Rezeptartige Standardlösungen sind daher nicht zu empfehlen. Dabei gilt es, die Balance zwischen Perturbation und Wertschätzung zu halten und das positive Selbstbild der Lernenden nicht generell zu gefährden, sowie ebenso als Lehrender die nun vielleicht deutlicher werdende Fehlbarkeit der eigenen Lehre auszuhalten.

Im Gegenzug zur vermeintlichen Gewissheit, dass das gesagte schon irgendwie verstanden wurde, entsteht aus der Kenntnis der Lernschwierigkeiten eine neue Ungewissheit: der Unterricht muss flexibel gestaltet werden, die Lenkungsrolle tritt etwas zurück, und ungewohnte Fragen können auch erfahrene Praktiker an ihre Grenzen bringen. Das Verständnis für die Schwierigkeiten der Lernenden wächst, während die Illusion der Wirksamkeit eigener, teilweise liebgewonnener Erklärungen abgebaut wird. All dies erfordert einen Lernprozess im Lehrenden, der aus den Feedbackelementen des PI und JiTT gespeist wird und zu einer Lehrhaltung der reflektierten Praxis führen kann - der Forschung zu und mit der eigenen Lehre. Hier wird die zweite Herausforderung von PI deutlich, welche dem Konzeptwandel bei Lernenden in ihrer Schwierigkeit sicher nicht nachsteht.

Selbst wenn alle Rahmenbedingungen optimal sind können schließlich auch die besten, mit umfangreichster Planung und Kenntnis typischer Fehlkonzepte entwickelten, Konzeptfragen noch immer kein Lernen garantieren! Feinheiten in der Umsetzung können starke Wirkung entfalten, weswegen einige Vorschläge zu einem sensibleren Umgang mit den Methoden angeboten wurden, die bei der möglichen Anpassung der Methode Impulse setzen könnten.

Das vorgestellte Konzept ist sicherlich kein nebenwirkungsfreies Allheilmittel und entspricht auch nicht dem Idealbild konstruktivistischer Lehre, kann aber in einigen Lernenden und Lehrenden Prozesse anregen, die einen Wandel in Richtung konstruktivistischer Lernprozesse in Gang setzen, ohne dabei zu abstrakt-geisteswissenschaftlich anzumuten. Gerade für Naturwissenschaftler/innen oder Ingenieure und Ingenieurinnen erleichtert dies den Einstieg. In diesem Wandel in der Einstellung zur eigenen Lehre und zum eigenen Lernen sehe ich den eigentlichen, verborgenen und sicherlich konstruktivistischen Nutzen der Methoden - die weitere Professionalisierung der Lehrenden und Lernenden.

# Literaturverzeichnis: